# Private Information, Credit Risk and Graph Structure in P2P Lending Networks

J. Christopher Westland: westland@uic.edu, University of Illinois – Chicago, USA

Tuan Q. Phan;   National University of Singapore; ;   email: tuan.quang.phan@gmail.com

Tianhui Tan; National University of Singapore; email: tinweitam@gmail.com

*Abstract*

This research investigated the potential for improving Peer-to-Peer (P2P) credit scoring by using "private information" about communications and travels of borrowers.  We found that P2P borrowers' ego networks exhibit scale-free behavior driven by underlying preferential attachment mechanisms that connect borrowers in a fashion that can be used to predict loan profitability.  The projection of these private networks onto networks of mobile phone communication and geographical locations from mobile phone GPS potentially give loan providers access to private information through graph and location metrics which we used to predict loan profitability.  Graph topology was found to be an important predictor of loan profitability, explaining over 5 ½% of variability.  Networks of borrower location information explain an additional 19% of the profitability.  Machine learning algorithms were applied to the data set previously analyzed to develop the predictive model and resulted in a 4% reduction in mean squared error.

## 1. The Peer-to-Peer Lending Industry

Peer-to-peer lending (P2P lending) involves the lending money to individuals or businesses through online services that match lenders with borrowers. The advantages of P2P lending arise from most business being conducted online; disadvantages arise from the perceived higher default risk than with loans made face-to-face, a perception that we investigated here. Many P2P loans are unsecured personal loans, though some of the largest amounts are lent to businesses; other forms of P2P lending include student loans, commercial and real estate loans, payday loans, as well as secured business loans, leasing, and factoring. The interest rates can be set by lenders who compete for the lowest rate on either a reverse auction or a rate fixed by an intermediary company based on an analysis of the borrower's credit. The lender's investment in the loan is not normally protected by any government guarantees. The lending intermediaries are for-profit businesses; they generate revenue by collecting a one-time fee on funded loans from borrowers and by assessing a loan service fee. Early P2P platforms had few restrictions on borrower eligibility, which resulted in adverse selection problems and high borrower default rates. In addition, because P2P loans typically have a minimum three-year term, some investors take a negative view of their lack of liquidity.

Privacy regulations, e.g., which guarantee the ability of an individual or group to seclude information about themselves, insert opportunities for moral hazard an adverse selection into the P2P lending business. (Böhme and Pötzsch 2010) asserted that the financial and social objectives of lending are inherently incompatible; (Grodzinsky and Tavani 2005) in a case study revealed exactly how those objectives might diverge; while (Dillon and Lending 2010) analyzed the reduced accuracy that accompanies improvements in privacy. In most developed economies of the world, financial privacy is rigorously protected by government. Lenders often complain that if they were only able to know more about their lenders' daily activities, communications, and so forth, that they could reduce the cost of lending, offer their customers better rates, and guarantee their investors more profitability. Sadly, governmental regulations have increased over time, and it seems unlikely that lenders will be given access to greater customer personal information in the near future. An entirely different situation exists in many developing economies, where urban or village life offers less opportunity for privacy, and borrowers are willing and legally allowed to surrender privacy of communication and travel in exchange for access to capital. In the current research, we have use this situation to our advantage to obtain a large collection of such data from a large consumer credit platform operating in a number of developing economies.

Developing economies have recently experienced intense innovation for P2P lending, credit scoring, risk analysis, customer vetting and collections, largely because regulation those countries allows the moderation of credit risk by giving lenders access to more private information from borrowers. China currently leads the world in new financial technology business models (Economist 2017). The company that provided data for the current analysis currently operates in 20 different countries throughout Asia. This company provides a ***software-*** *as-a-service* platform relying on non-traditional data from social media and smartphone records in order to ascertain customers' financial stability. Its vision is stated to be "to improve financial inclusion for at least a billion people" in developing economies around the world. They partner with traditional and alternative lenders in providing loans.

Our current research expands the tools available for credit risk analysis by asking the research question:

> "Can graph theoretic models of credit default risk with access to "private information" about borrower communication and travel increase loan interest revenues less default costs from current "best practice"

Micro-level studies that focus on the individual in a social setting are referred to as "ego network" analyses, where the focus is individual nodes or "actors" in the network (Everett and Borgatti 2005, Arnaboldi, Conti et al. 2012, Leskovec and Mcauley 2012). At the macro level, graph metrics may not directly assess factors in prevention, credit scoring, costs and collection outcomes; rather they estimate epiphenomena. Epiphenomena are secondary phenomenon that occur alongside or in parallel to measurable causal phenomena; they arise from, but do not directly influence a process. Signs, symptoms, warnings and other credit risk factors can all be epiphenomena in this sense. Graph theoretic epiphenomena hold the potential to predict emergent health behavior and outcomes unavailable to analytics that fail to consider such epiphenomena.

The current research investigates causal factors in overall lending profitability by testing seven hypotheses supporting investigation of the research question of whether net loan interest revenues less loan default costs (our measure of contribution to profit) are predicted by [specific borrower metrics]

$H_1$ (Naïve − Chartist): Future profitability depends only on past profitability

$H_2$ (Baseline): Future profitability is predicted by past defaults, interest rates and principal

$H_3$ (+Graph Topology): Future profitability is predicted by past defaults, interest rates, principal and borrower communications graph metrics

$H_4$ (+Location) : Future profitability is predicted by past defaults, interest rates, principal, borrower communications graph metrics and borrower geographic proximity to particular classes of business

$H_5$ (+Machine Learning ): Future profitability is best predicted by computationally intensive machine learning algorithms that base their decisions on past defaults, interest rates, principal, borrower communications graph metrics and borrower geographic proximity to particular classes of business.

$H_6$ : P2P borrowers' ego networks exhibit scale-free behavior driven by underlying preferential attachment mechanisms that connect borrowers in a fashion that potentially could be used to predict credit defaults.

*Importance*: Prior research has suggested that scale-free behavior is common in ego networks. The scale-free property is significant because it allows us to analyze subgraphs, or even individuals in a population and comfortably infer that the population has similar characteristics.

These six hypotheses $H_1$ to $H_6$ will be tested against our dataset. Our investigation proceeds as follows. Section 2 reviews prior literature that has influenced current best-practices in P2P landing; section 3 details the data and generation of graph models; section 4 tests for confounding influences; section 5 tests the predictors of P2P lending profitability; section 6 builds and tests our scale-free network structural model; section 7 discusses the findings in the context of current lending practice.

## 2. Prior Literature
*Private Information*

Many legal systems prohibit opportunities for lenders to expanding liquidity and access to credit due to fear of being accused of "digital redlining" where the eligibility of a person to get a loan is decided by algorithms. The Electronic Privacy Information Center suggests there is a "general risk caused by using personal information to clarify and sort people and determine who is a high-value consumer and who is not worth engaging with." Their "general risk" is not further specified; but where credit decisions must be made, one can argue that with more information and greater transparency lenders will be more, rather than less, likely to contract with borrowers. (Mother Jones "Your Deadbeat Facebook Friends Could Cost You a Loan" September 2013). The provider of our dataset has made this argument, citing its particular relevance to emerging economies, where access to credit takes precedence over information privacy concerns.

Much of the information of significance to credit scoring resides in the structure of social networks are mined to extract information that can improve liquidity and access.

At the micro-level, social network research typically begins with an individual, snowballing as social relationships are traced, or may begin with a small group of individuals in a particular social context (Wasserman and Faust 1994, Scott and Davis 2003, Scott 2017). Micro level analysis may occur at the level of the dyad a social relationship between two individuals where the focus analysis is the structure of the pairwise relationship and tendencies toward reciprocity/mutuality; or the triad where they concentrate on balance and transitivity (Kadushin 2012). Studies that focus on the individual in a social setting are called "ego networks," where the focus is individual nodes or "actors" in the network (Everett and Borgatti 2005, Arnaboldi, Conti et al. 2012, Leskovec and Mcauley 2012). They focus on characteristics such as size, relationship strength, density, centrality, prestige and roles such as isolates, liaisons, and bridges (Jones and Volpe 2011) and are used in the fields of psychology, ethnography and genealogy (de Nooy 2012).

At the macro-level, analyses trace the outcomes of interactions, such as economic resource transfers amongst a population. Macro-level social networks display features of social complexity, which involves substantial non-trivial features of network topology, with patterns of complex connections between elements that are neither purely regular nor purely random and are distinguished by a heavy tailed degree distribution, a high clustering coefficient and community structure (Callaway, Newman et al. 2000, Strogatz 2001, Borgatti, Mehra et al. 2009, Easley and Kleinberg 2010).

## 3. Data and Graph Generation

The current research is conducted on a large dataset of mobile phone communications for a subset of 784 borrowers, obtained from a P2P lending company which operates in 20 different countries. The set of loans used in this research were extended in a country where privacy laws are relatively lax, and the P2P lending company is allowed access to complete phone records of borrowers under contract. The data set combines SMS (short message services) communications with voice communications for each handset. SMS communications are by default given an average communications duration of one minute in this research, which was an estimate of the equivalent amount of information that would be communicated by voice. Call duration was interpreted as a surrogate for information conveyed in a communication, with more information indicating a "stronger" relationship between a caller and receiver. 19.39% of the 784 loans were in default, representing 21.73% of loaned value ($2346040 / $10795722) in our dataset.

Our raw data contained 4,142,474 individual SMS and voice communications. We curated these 4,142,474 individual communications using a variety of our language data cleaning functions, removing informational or emergency calls; standardizing all caller identifiers into 11-digit telephone numbers used in the country; and eliminating error and nonsense identifiers. The resulting 3,577,912 caller identifiers were aggregated on caller-receiver dyads summing over the call durations to generate an edge-link list where each link is total quantity of exchange of 'information' (call durations) between individuals.

*Table 1: Research Dataset 1.5 years of call data from Jan 1, 2014 to June 27, 2015. Loan origination dates run from 2012-12-15 to 2014-08-11. Calls originated between 0 days after the loan origination to 366 days after the loan origination.*

| Dataset | Count | Descriptors |
| --- | --- | --- |
| *Loans* | 784 | " borrower"; "default"; "amount"; "interest"; "contracting time" |
| *Voice Communications by borrowers* | 564,562 | "borrower"; " call to/from"; "type"; "call time"; "duration" |
| *SMS Communications by borrowers* | 3,577,912 | "borrower"; " SMS to/from"; "type"; "SMS time"; "duration=1" |
| *Businesses within a 50-meter radius at time of communication by borrowers (obtained by comparing latitude-longitude with businesses listed on Google maps)* | 81,050 | " borrower"; "longitude"; "latitude"; "call time"; "time since loan contract"; ["set of 107 Google Maps business types (counts in 50 meter radius of phone)"] |
| *Communication graph size (vertices) after aggregation on duration* | 112,487 | " borrower"; "default (blue = default)"; "'information' (edge weight)" |

*Figure 1: Hourly Location of Borrower Calls on Four Maps of Increasing Scale (Colors run from Red through Purple / Magenta representing Midnight to Midnight*

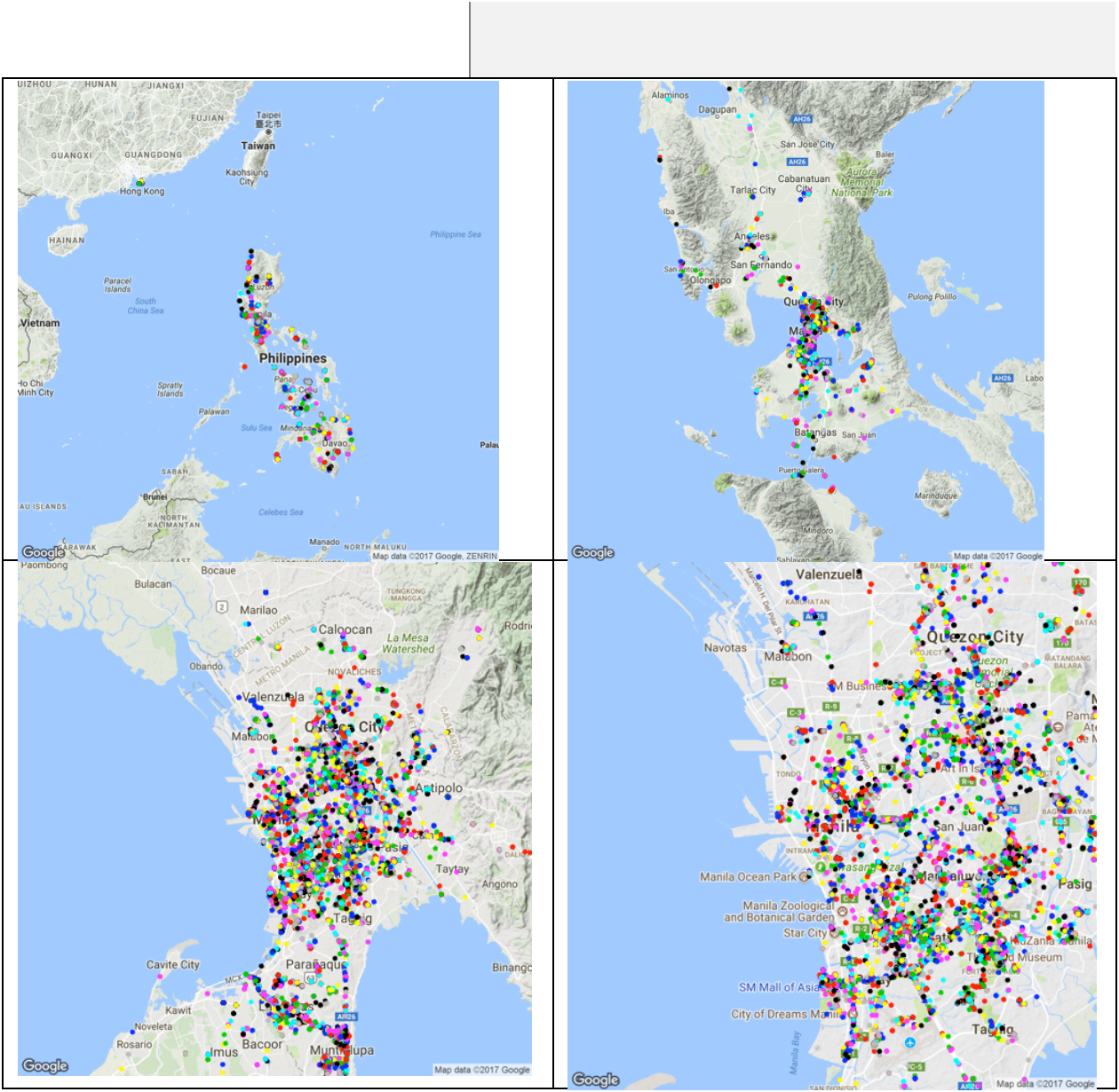

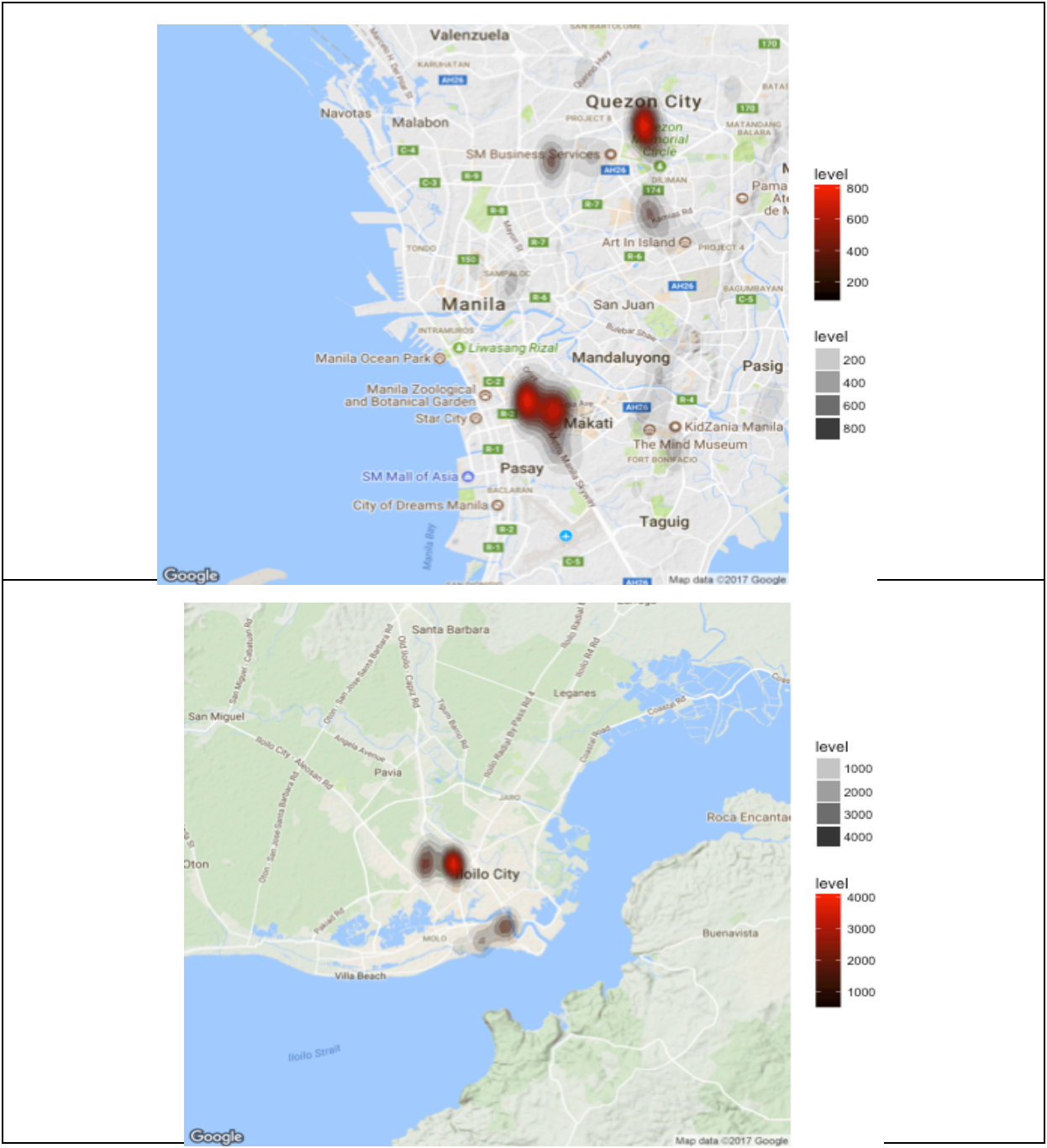

*Figure 2: Heat Map of Location of Heaviest Mobile Usage by Borrowers (Manila and Iliolo City)*

Hypotheses 1 and 2 were tested using the full dataset of $784 \ltimes 3{,}577{,}912$ left semi-joins of loans and communications. The dataset was used to create directed graphs wherever $n > 2$ communications occurred, edges were defined for the $\binom{n}{2} = \frac{n!}{2(n-2)!}$ combinations of nodes.

Graph construction data cleaning and plotting we're accomplished using the *igraph* package in R (Csardi and Nepusz 2006, Csárdi and Nepusz 2010). Graph statistics for centralization, degree, modularity, community structure, diameter, connectedness and other statistics were computed using SNAP (Leskovec and Sosič 2016). The comprehensive graph was used to analyze the empirical graph structure and the influence of an individual borrower using whole graph metrics. Default risks were analyzed using general linear models of communication, location and loan contract variables using appropriate residual distributions and link functions.

## 5. Predictors of Loan Profitability

Often loan credit scoring is seen as a process of dichotomizing borrowers into good borrowers or borrowers who are likely to default. From a business standpoint, this is not as useful as identifying borrowers who are expected to be profitable for the P2P lending firm. The P2P lending firm may be willing to take on risky borrowers at higher interest rates or smaller principal, and indeed we perceive this in practice. Loan profitability is the net of loan revenue less loan expenses. Annual loan revenue is interest multiplied by principal; expenses derived from defaults, where the actual outcome of a default is complex. Borrowers may default on the entire principal or only a portion of it, or only miss an interest payment. Once the borrower is in default the loan may be rescheduled, or the collateral may be sold and used to offset the loss from the loan default.

In the current research, we did not have complete information to fully characterize the loans on the database. For example, we did not have the term of the loan, or the collateral pledged to secure the loan, or information on collections after default. We could construct a variable for the amount of time that had passed since the loan contract was written and the last communication was registered in our data set; but this proved to be a very noisy and imprecise metric. We therefore constructed the following surrogate to track loan profitability:

$$\text{synthetic } annual \text{ profitability} = \begin{cases} principal \times interest & if\ good\ loan \\ 0 & if\quad default \end{cases}$$

Our major parallels the methods used in practice by large credit service bureaus in credit scoring. Interest rates and amount principle are connected to borrower default rates as well as interest rates and the profitability of the leading company. Actual practice is not just focused on default but also on the overall profitability of performing loans. Obviously, the best situation for the lender is to have many performing loans with very high interest rates and large principal amounts. Conversely, high interest rates are associated with the risky borrowers, thus any credit decision needs to consider default probability as well as interest charged and principal loaned. Our synthetic profitability contains all the information in principal, interest and defaults to provide the best decision metric for profitability using our predictors on the database. Figure 4 graphs the histogram of synthetic profitability for the 784 loans on the dataset.

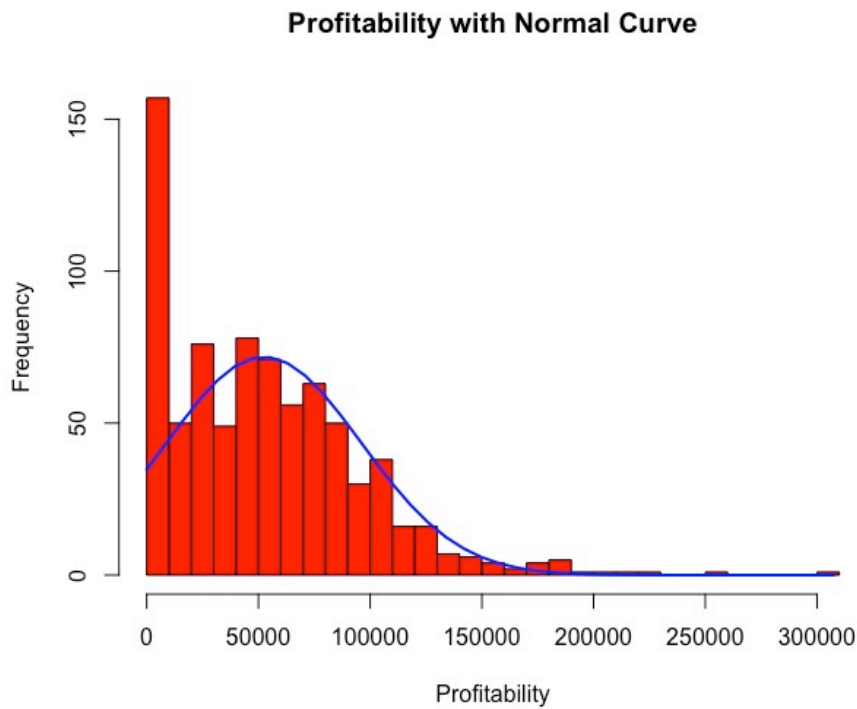

*Figure 3: Distribution of synthetic profitability with a Normal density curve fitted to that data*

Our synthetic profitability metric is zero inflated with approximately 19.4% of loans having a value of zero. This suggests that we should fit the data using the zero an inflated distribution. The most commonly used zero inflated distribution is the zero inflated Poisson which is used to analyze count data in the insurance industry. The zero-inflated Poisson model employs two components that correspond to two zero generating processes. The first process is governed by a binary distribution that generates structural zeros. The second process is governed by a Poisson distribution that generates counts, some of which may be zero.

Table 6 reports the results of a regression assuming a Tobit distribution and uses the Vuong test to compare to ordinary least squares with normal errors. The Normal distribution is the continuous counterpart to the Poisson distribution and is more suitable for our do dataset. The tests in tables 5 and 6 confirm that despite zero inflation of our synthetic profitability, zero inflation is small enough at 19.4% that without loss of generality, we can safely use ordinary least squares regression for our analysis in the remainder of the paper.

*Table 2: Test of Zero-Inflated Poisson Distribution (model = profit ~ 1)*

|  | Estimate | s.e. |
|---|---|---|
| Count model coefficients (Poisson with log link): | 6.47732 | 0.00156 |
| Zero-inflation model coefficients (binomial with logit link): | -1.42501 | 0.09034 |
| Log-likelihood: | -71870 on 2 df | |
| Vuong Non-Nested Hypothesis z-Statistic: | 13.97480; p-value = 0.0000; Model is indistinguishable from GLM Poisson with log link | |

*Table 3: Test of Zero-Inflated Tobit Distribution (model = profit ~ 1)*

|  | Estimate | Std. | z | $Pr(>|z|)$ | |
|---|---|---|---|---|---|
| (Intercept) | 46940 | 1912.00 | 24.55 | 0.0000 | *** |
| Log(scale) | 11 | 0.03 | 367.73 | 0.0000 | *** |
| Log-likelihood: | -7919 | | | | > |
| Vuong Non-Nested Hypothesis z-Statistic: | 15.390; p-value = 0.0000; Model is indistinguishable from OLS Normal | | | | |

Table 6 reports the results of a regression assuming a Tobit distribution and uses the Vuong test to compare to ordinary least squares with normal errors. The Normal distribution is the continuous counterpart to the Poisson distribution and is more suitable for our do dataset. The tests in tables 5 and 6 confirm that despite zero inflation of our synthetic profitability, zero inflation is small enough at 19.4% that without loss of generality, we can safely use ordinary least squares regression for our analysis in the remainder of the paper.

Additionally, the timing of communications appeared not to have an impact on profitability. The *diff_day* predictor (mean = 212.3) is the difference between the time of communication and the time of the original loan contract. It appears that loan profitability is not heavily dependent on time or the timing of communications, as the estimator value of -3.5 suggests that the influence of time never contributes or subtracts more than about $1000 dollars from profitability.

We analyzed the occurrence of default and impact on profitability our set of loan predictors starting with naïve and baseline models; adding graph statistics for the communications of borrowers; and finally adding geographical location of borrowers and their proximity to particular types of businesses and organizations. We also tested regression based machine learning models to see if these can improve on the existing statistical regression approaches. We investigate the additional information provided by each cluster statistics with the intention of suggesting where access to phone records provides the greatest amount of information relevant to loan credit scoring. We constructed five models, paralleling the five nested hypotheses under $H_3$, to assess the relative information content of various predictors.

We constructed five models:

$H_1$ (Naïve − Chartist): Future profitability depends only on past profitability

$H_2$ (Baseline): Future profitability is predicted by past defaults, interest rates and principal

$H_3$ (+Graph Topology): Future profitability is predicted by past defaults, interest rates, principal and borrower communications graph metrics

$H_4$ (+Location) : Future profitability is predicted by past defaults, interest rates, principal,

borrower communications graph metrics and borrower geographic proximity to particular classes of business

$H_5$ (+Machine Learning ): Future profitability is best predicted by computationally intensive machine learning algorithms that base their decisions on past defaults, interest rates, principal, borrower communications graph metrics and borrower geographic proximity to particular classes of business.

| *hypothesis* | *Adj $R^2$* | *predictors* |
|---|---|---|
| $H_1$ | 0.0000 | $1s\ vector$ (naïve) |
| $H_3$ | 0.0554 | graph |
| $H_4$ | 0.1888 | location |
| $H_3$ & $H_4$ | 0.2243 | location +graph |
| $H_2$ | 0.8210 | loan |
| $H_2$ & $H_3$ & $H_4$ | 0.8625 | loan +graph +location |

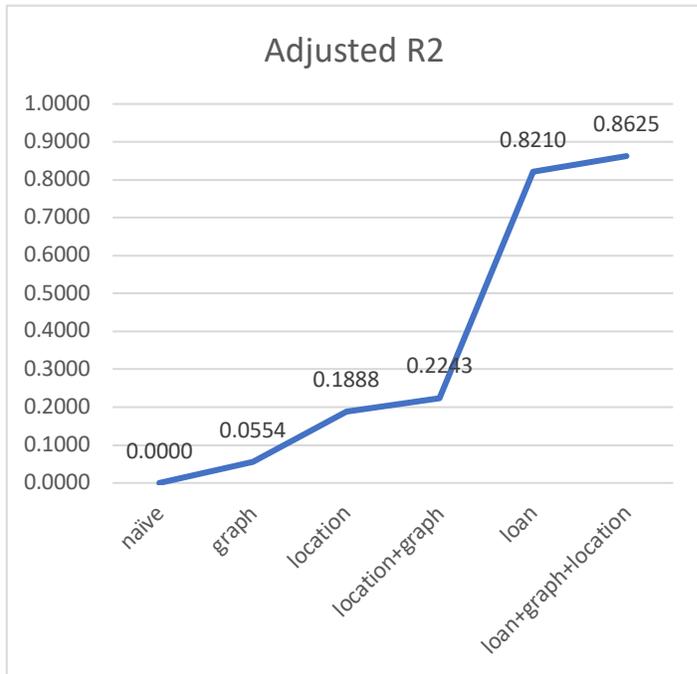

*Figure 4: Variance Explained*

The naïve model in the baseline model serve as benchmarks, presenting the situation where we use none of the information that was obtained in this research dataset. The naïve model is significant because there are technical models of prediction in financial markets that look only at prior history of an effect variable, for example a stock price.

The baseline model on the other hand, is designed to reveal how much loan information in the original data set is retained in our synthetic profitability measure. More than 82% of the original information is retained in our synthetic profitability measure, which implies that "synthetic profitability" is a much more informative alternative for credit scoring to the traditional dichotomous "loan default" indicator that is used in many machines learning studies.

Two important sets of information were elicited from the lender's database: (1) the graph topology of the network relationships formed by borrowers and their acquaintances exchanging SMSs and telephone calls; and (2) the locations visited by the borrowers, and types of businesses or institutions that might be close to those locations. The graph topology oh communication networks of borrowers explained an additional 5 ½% of the profitability variance; our information about types of businesses or institutions that might be close to borrower locations explain almost an additional 19% of the profitability variance. Together these explained around 22 ½% of the profitability variance, or $92,229.21 of the total profit (synthetic) of $411,186.80 for the 784 loans in our dataset. This sets a potential upper limit for the improvement in predictive performance graph and location information.

We could potentially combine this additional information with our baseline model as as reported in the rightmost score on figure 5; but this muddies the contribution graph topology and location information contribute to better credit scoring.

Our final nested hypothesis $H_5$ looks at the potential for machine learning based feature selection, cross-validation, and predictive analytics to improve on the statistical regression models used in the research up to this point.

Machine learning is usually invoked to predict future responses based on past observations with known response values. Linear regression models provide a benchmark for many machine learning methods, where the specification search problem is typically called "feature selection." Feature selection must address two conflicting concerns: (1) prediction accuracy, determined by the variance of the prediction error, and (2) prediction bias when applying the model results to new observations. As more predictors are introduced to a model, model accuracy usually improves, but bias may degrade because of noisy data, redundant predictors, multi-collinearity, missing values, outliers and so forth. Several approaches automate a standard statistical framework for feature selection and dimensionality reduction – stepwise regression, simulated annealing, principal components analysis and radial basis functions. Stepwise regression is most appropriate in our situation where we need to incorporate predictors' effects into the model. A variety of penalty functions can be used for feature selection in stepwise regression; prior modeling in this paper has focused on information content, thus we propose using the Akaiki information criterion (AIC).

Measures of information content are of fundamental importance in establishing the value of the new metrics proposed for credit scoring in this research. There are many different information metrics and these tend to measure roughly the same things, but in different contexts; we used several in the current research which we describe here. Fisher information is the amount of information that an observable random variable $x$ carries about an unknown parameter $\theta$ of a distribution of $x$. Fisher information is the variance of the score, i.e., the expected value of the observed information $(x_1, \ldots, x_n)$. The concept of distribution can be extended to that of a model $M$ for vector $x$ where maximum likelihood is $\hat{L} = \max_\theta L = \max_\theta (P(x|\theta, M))$. Where we estimate k-dimensional $\theta$ we can define the Akaike information criterion (Akaike 1973) $AIC = 2k - 2ln\hat{L}$. In the context of linear regression with a residual sum of squares (RSS), $AIC = 2k - nln(RSS)$. AIC is commonly used where models are compared, and in this context, the relative likelihood of two models $M_1$ and $M_2$ is a function of the AICs of the models $e^{\frac{AIC(M_1) - AIC(M_2)}{2}}$ (Burnham and Anderson 2003). Leave-one-out cross-validation is asymptotically equivalent to the AIC for ordinary linear regression models as well as for mixed-effects models (Fang 2011).

Table 4: Synthetic Profit ~ All Predictors (OLS); R-squared: 0.863;  F: 4530 on 103 and 74177 DF; p-value:0.00000

| Change secret this | Estimate | Pr(>\|t\|) |  | Predictor | Estima | Pr(>\|t\|) |  | Predictor | Estima | Pr(>\|t\|) |
|---|---|---|---|---|---|---|---|---|---|---|
| (Intercept) | 7110 | 0.00 | *** | amt | 3.1 | 0.00 | *** | taxi_stand | -13500 | 0.05 |
| art_gallery | 7800 | 0.00 | *** | far | 2.9 | 0.00 | *** | bar | -764 | 0.05 |
| eigen | 1060000000 | 0.00 | *** | diff_day | -3.6 | 0.00 | *** | department | 796 | 0.05 |
| dentist | 13800 | 0.00 | *** | finance | -1020 | 0.00 | *** | storage | 5260 | 0.08 |
| pharmacy | 10100 | 0.00 | *** | school | -1070 | 0.00 | *** | lawyer | 656 | 0.10 |
| car_rental | 10000 | 0.00 | *** | store | -1110 | 0.00 | *** | car_repair | 270 | 0.12 |
| funeral_home | 9400 | 0.00 | *** | lodging | -1180 | 0.00 | *** | liquor_store | 2300 | 0.12 |
| veterinary_care | 7940 | 0.00 | *** | grocery_or_superm | -1310 | 0.00 | *** | hair_care | 651 | 0.13 |
| car_wash | 7620 | 0.00 | *** | meal_delivery | -1690 | 0.00 | *** | real_estate_ag | -336 | 0.13 |
| post_office | 6940 | 0.00 | *** | general_contractor | -1790 | 0.00 | *** | painter | -27900 | 0.15 |
| int | 6860 | 0.00 | *** | insurance_agency | -2280 | 0.00 | *** | home_goods_s | 375 | 0.17 |
| furniture_store | 6700 | 0.00 | *** | car_dealer | -3160 | 0.00 | *** | plumber | -8100 | 0.17 |
| embassy | 5890 | 0.00 | *** | gas_station | -3220 | 0.00 | *** | cemetery | 4470 | 0.19 |
| library | 5480 | 0.00 | *** | ins | -5640 | 0.00 | *** | parking | 1350 | 0.20 |
| museum | 5090 | 0.00 | *** | health | -5650 | 0.00 | *** | electrician | -8750 | 0.21 |
| pet_store | 4850 | 0.00 | *** | neighborhood | -5740 | 0.00 | *** | clothing_store | 372 | 0.24 |
| natural_feature | 4770 | 0.00 | *** | night_club | -7630 | 0.00 | *** | synagogue | -23000 | 0.25 |
| doctor | 4630 | 0.00 | *** | bus_station | -8490 | 0.00 | *** | bicycle_store | 2420 | 0.31 |
| travel_agency | 4590 | 0.00 | *** | florist | -9320 | 0.00 | *** | book_store | -609 | 0.34 |
| local_government_ | 4440 | 0.00 | *** | out | -9400 | 0.00 | *** | mosque | 12600 | 0.36 |
| spa | 4150 | 0.00 | *** | church | -11700 | 0.00 | *** | bowling_alley | -4200 | 0.37 |
| electronics_store | 3790 | 0.00 | *** | airport | -13200 | 0.00 | *** | fire_station | -1350 | 0.42 |
| shoe_store | 3780 | 0.00 | *** | city_hall | -21100 | 0.00 | *** | casino | 578 | 0.46 |
| gym | 3400 | 0.00 | *** | def | - | 0.00 | *** | campground | -4160 | 0.47 |
| movie_theater | 3220 | 0.00 | *** | beauty_salon | -1060 | 0.00 | ** | dur | -0.1 | 0.51 |
| shopping_mall | 3140 | 0.00 | *** | train_station | -6640 | 0.01 | ** | light_rail_statio | -1160 | 0.56 |
| jewelry_store | 2780 | 0.00 | *** | bakery | -685 | 0.01 | ** | locksmith | 2080 | 0.58 |
| laundry | 2450 | 0.00 | *** | restaurant | 383 | 0.01 | * | convenience_st | -213 | 0.58 |
| hospital | 2150 | 0.00 | *** | moving_company | -9100 | 0.01 | * | accounting | -351 | 0.59 |
| university | 1150 | 0.00 | *** | place_of_worship | 4240 | 0.02 | * | stadium | 5920 | 0.60 |
| cafe | 943 | 0.00 | *** | transit_station | 3830 | 0.02 | * | roofing_contrac | 4200 | 0.71 |
| meal_takeaway | 882 | 0.00 | *** | police | -3570 | 0.02 | * | park | -395 | 0.76 |
| food | 834 | 0.00 | *** | hardware_store | -989 | 0.02 | * | atm | 83.1 | 0.78 |
| triad | 262 | 0.00 | *** | movie_rental | -30500 | 0.03 | * | bank | -56.3 | 0.83 |
|  |  |  |  | amusement_park | 11400 | 0.04 | * | aquarium | 1510 | 0.94 |
|  |  |  |  |  |  |  |  | hindu_temple | NA | NA |

*Table 5: Feature selection where Synthetic Profit ~ SWA AIC-max Predictors (OLS); R-squared:0.863; F-statistic:5980 on 78 and 74202 DF, p-value: 0.00000*

| Predictor | Estimate | Pr(>\|t\|) |  | Predictor | Estimate | Pr(>\|t\|) |  | Predictor | Estimate | Pr(>\|t\|) |  |
|---|---|---|---|---|---|---|---|---|---|---|---|
| (Intercept) | 7140 | 0.00 | *** | jewelry_store | 2810 | 0.00 | *** | florist | -9320 | 0.00 | *** |
| eigen | 1070000000 | 0.00 | *** | laundry | 2510 | 0.00 | *** | out | -9550 | 0.00 | *** |
| dentist | 13900 | 0.00 | *** | hospital | 2150 | 0.00 | *** | church | -11500 | 0.00 | *** |
| car_rental | 10100 | 0.00 | *** | university | 1140 | 0.00 | *** | airport | -13200 | 0.00 | *** |
| pharmacy | 10000 | 0.00 | *** | cafe | 1030 | 0.00 | *** | city_hall | -21200 | 0.00 | *** |
| funeral_home | 9620 | 0.00 | *** | meal_takeaway | 912 | 0.00 | *** | def | -103000 | 0.00 | *** |
| car_wash | 8080 | 0.00 | *** | food | 752 | 0.00 | *** | transit_station | 3310 | 0.01 | ** |
| veterinary_care | 7820 | 0.00 | *** | restaurant | 464 | 0.00 | *** | bakery | -742 | 0.00 | ** |
| art_gallery | 7700 | 0.00 | *** | triad | 262 | 0.00 | *** | beauty_salon | -1070 | 0.00 | ** |
| furniture_store | 6970 | 0.00 | *** | amt | 3.1 | 0.00 | *** | train_station | -6430 | 0.01 | ** |
| post_office | 6920 | 0.00 | *** | far | 2.9 | 0.00 | *** | moving_comp | -9270 | 0.01 | * |
| int | 6860 | 0.00 | *** | diff_day | -3.5 | 0.00 | *** | place_of_wors | 4080 | 0.02 | * |
| embassy | 5900 | 0.00 | *** | school | -1010 | 0.00 | *** | hardware_sto | -997 | 0.02 | * |
| library | 5460 | 0.00 | *** | finance | -1040 | 0.00 | *** | police | -3600 | 0.02 | * |
| museum | 5060 | 0.00 | *** | store | -1040 | 0.00 | *** | movie_rental | -31000 | 0.02 | * |
| pet_store | 4830 | 0.00 | *** | lodging | -1200 | 0.00 | *** | amusement_p | 11600 | 0.04 | * |
| doctor | 4660 | 0.00 | *** | grocery_or_super | -1300 | 0.00 | *** | taxi_stand | -13400 | 0.05 | . |
| travel_agency | 4660 | 0.00 | *** | meal_delivery | -1580 | 0.00 | *** | bar | -732 | 0.06 | . |
| natural_feature | 4620 | 0.00 | *** | general_contracto | -1790 | 0.00 | *** | storage | 5410 | 0.07 | . |
| local_government | 4380 | 0.00 | *** | insurance_agency | -2300 | 0.00 | *** | department_s | 674 | 0.09 | . |
| spa | 4150 | 0.00 | *** | gas_station | -3200 | 0.00 | *** | real_estate_a | -369 | 0.10 | . |
| shoe_store | 4030 | 0.00 | *** | car_dealer | -3270 | 0.00 | *** | liquor_store | 2370 | 0.11 |  |
| electronics_store | 3830 | 0.00 | *** | health | -5680 | 0.00 | *** | lawyer | 627 | 0.11 |  |
| gym | 3430 | 0.00 | *** | ins | -5700 | 0.00 | *** | plumber | -9110 | 0.12 |  |
| shopping_mall | 3060 | 0.00 | *** | neighborhood | -5800 | 0.00 | *** | hair_care | 647 | 0.12 |  |
| movie_theater | 3050 | 0.00 | *** | night_club | -7610 | 0.00 | *** | painter | -28100 | 0.15 |  |
|  |  |  |  | bus_station | -7990 | 0.00 | *** |  |  |  |  |

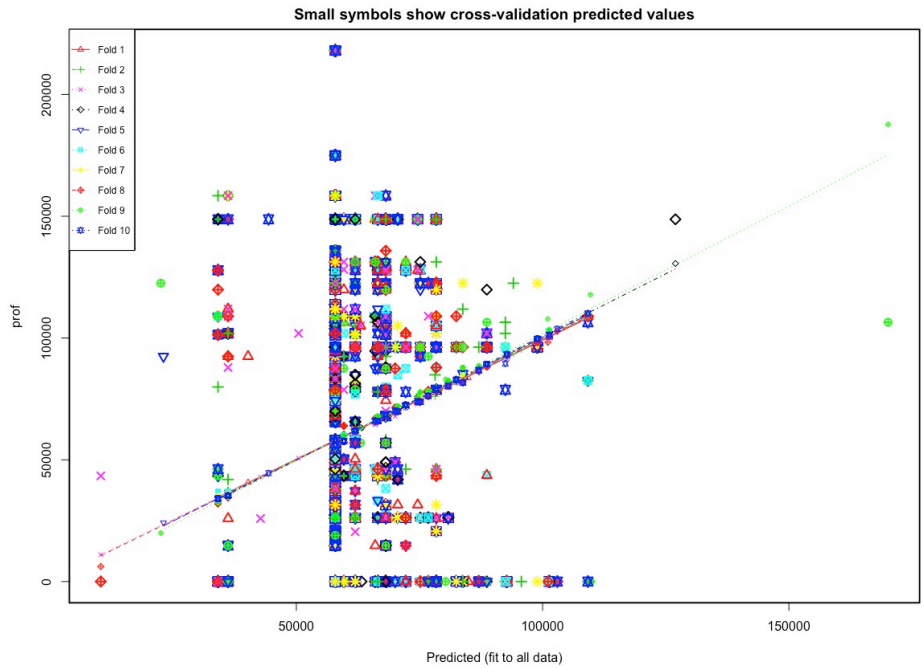

*Figure 5: 10–fold cross validation of the original model (synthetic profit ~ loan +graph +location predictors) ;MSE = 3.94e+08*

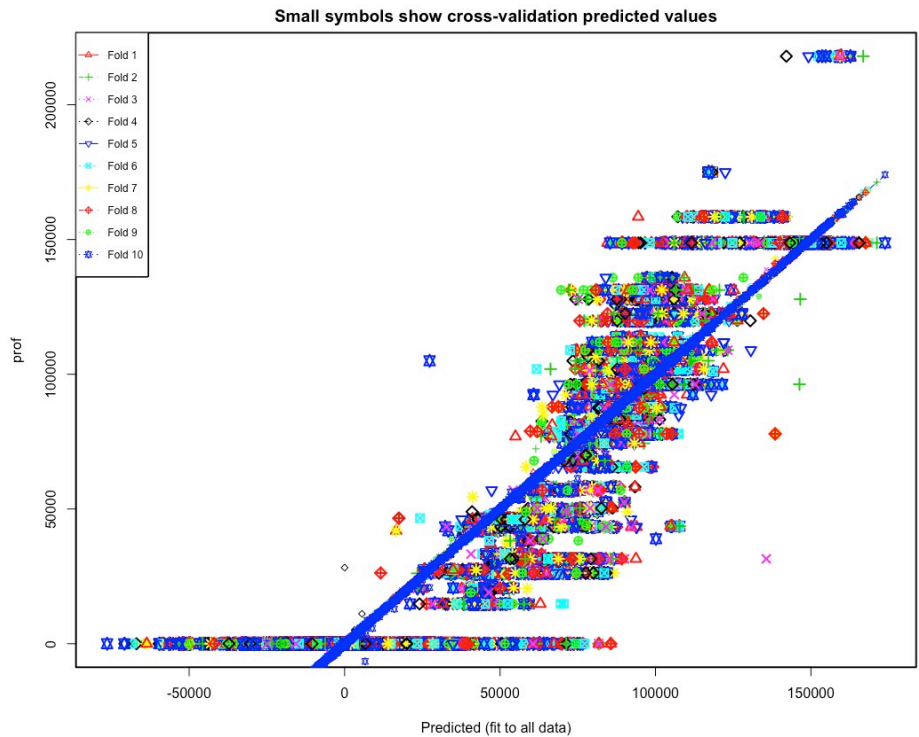

*Figure 6: 10–fold cross validation of the AIC-maximizing feature set of predictors; MSE = 3.79e+08*

Our machine learning investigation picks up where we left off with the nested linear regression models – the model using the full data set (synthetic profit ~ loan +graph +location predictors). Estimators and p-values from this model are summarized in table 7. We then applied a machine learning stepwise regression algorithm with an AIC penalty function to select information rich features from the synthetic profit ~ loan +graph +location predictors model. Machine learning feature selection produced the bottle in table 8, which for the most part removes the predictors with the least significant t-scores.

$H_1$ (Naïve − Chartist): Future profitability depends only on past profitability

$H_2$ (Baseline): Future profitability is predicted by past defaults, interest rates and principal

$H_3$ (+Graph Topology): Future profitability is predicted by past defaults, interest rates, principal and borrower communications graph metrics

$H_4$ (+Location) : Future profitability is predicted by past defaults, interest rates, principal, borrower communications graph metrics and borrower geographic proximity to particular classes of business

$H_5$ (+Machine Learning ): Future profitability is best predicted by computationally intensive machine learning algorithms that base their decisions on past defaults, interest rates, principal, borrower communications graph metrics and borrower geographic proximity to particular classes of business.

At this point in the research we can draw conclusions about the research support or rejection are nested hypotheses $H_3$ to assess the relative information content of various predictors.

$H_1$ (Naïve − Chartist): $H_{3a}$ (Naïve − Chartist): our research concluded that profitability and series alone are insufficient to predict future along profitability. The notion of there being a chartist approach to loan credit scoring was always a bit far-fetched, given the zero inflated distribution of the parameters. The naïve model is significant because there are technical models of prediction in financial markets that look only at prior history of an effect variable, for example a stock price.

$H_2$ (Baseline): $H_{3b}$ (Baseline): The baseline model reveals how much loan information in the original data set is retained in our synthetic profitability measure. More than 82% of the original information is retained in our synthetic profitability measure, which implies that "synthetic profitability" is a much more informative alternative for credit scoring to the traditional dichotomous "loan default" indicator that is used in many machines learning studies.

| Graph metric | Estimator | Mean value of data |
|---|---|---|
| Number of triads containing this borrower | 262 | 6.94308 |
| Farness centrality | 2.9 | 37.76445 |
| Eigenvector centrality | 1070000000 | 0.00260 |
| Number of outgoing edges | -9550 | 292.54590 |
| Number of incoming edges | -5700 | 0.70820 |

Table 6: Graph metric predictors, all values are significant at the 99.99% level

$H_4$ (+Location) : $H_{3d}$ (+Location) Tables 7 and 8 were sorted by predictor p-values and then by estimator values, which emphasize the most influential predictors for our synthetic profit measure. Table 10 summarizes the most influential predictors on loan profitability and their impact on loan profitability.

| locations that increase | net$ influence of 1 | locations that decrease | net$ influence of 1 |
|---|---|---|---|
| dentist | 13900 | city hall | -21200 |
| car rental | 10100 | airport | -13200 |
| pharmacy | 10000 | church | -11500 |
| funeral home | 9620 | florist | -9320 |
| car wash | 8080 | bus station | -7990 |
| veterinary care | 7820 | night club | -7610 |
| art gallery | 7700 | neighborhood | -5800 |
| furniture store | 6970 | health | -5680 |
| post office | 6920 | car dealer | -3270 |
| embassy | 5900 | gas station | -3200 |
| library | 5460 | insurance agency | -2300 |
| museum | 5060 | general contractor | -1790 |
| pet store | 4830 | meal delivery | -1580 |
| doctor | 4660 | grocery or supermarket | -1300 |
| travel agency | 4660 | lodging | -1200 |
| natural feature | 4620 | finance | -1040 |
| local government office | 4380 | store | -1040 |
| spa | 4150 | school | -1010 |
| shoe store | 4030 | | |
| electronics store | 3830 | | |
| gym | 3430 | | |
| shopping mall | 3060 | | |
| movie theater | 3050 | | |
| jewelry store | 2810 | | |
| laundry | 2510 | | |
| hospital | 2150 | | |
| university | 1140 | | |
| cafe | 1030 | | |

*Table 7: Location predictors: businesses that add or subtract more than $1000 from synthetic profitability if located within 50 m of borrower*

$H_3$ (+Graph Topology): $H_{3c}$ (Graph Topology) Graph topology was found to be an important predictor of loan profitability, explaining over 5 ½% of variability in our synthetic profitability measure, or $22,615.27 of the total profit (synthetic) of $411,186.80 for the 784 loans in our dataset. Table 9 presents these results. One problem in interpreting this data is that the scale of the graph metrics varies substantially between metrics; so it is important to analyze both the estimated value and the average value of the data item. Table 9 indicates that the greater centrality of a particular borrower in the communication network is correlated strongly with loan profitability; while high numbers of outgoing edges (communications to others) correlate strongly to losses and defaults.

Our location data is represented as count data; it is the number of businesses of a particular classification that are within 50 m of the borrower at a particular communication time. Although any observations are inherently subjective, we can surmise that a proximity of certain types of businesses reflects the borrowers' particular lifestyle choices, and these lifestyle choices impact ability to repay loans on time. Locations such as dentist offices, pharmacies, veterinary care facilities, libraries and so forth reflect lifestyles of borrowers who take responsibility for their homes and families pass and probably loans. Locations such as nightclubs, bus stations,

City Hall, airports and financial offices may reflect lifestyles of people who are more carefree, and perhaps more likely to get into financial trouble.

Location information about types of businesses or institutions that where we been 50 m of the borrower at the time of communication explain an additional 19% of the synthetic profitability variance, or $77,632.07 of the total profit (synthetic) of $411,186.80 for the 784 loans in our dataset.

Another way to determine the influence of location graph network related artifacts on profitability is to analyze the Cook's distance of each observation (Cook 1977, Cook 1979). Cook's distance commonly used to estimate the influence of a data point when performing a least-squares regression analysis and can be interpreted as the distance one's estimates move within the confidence ellipsoid that represents a region of plausible values for the parameters. In descending sequence, table 11 shows the predictors that should have the greatest influence on our model's forecast of profitability.

*Table 8: Predictors with the greatest influence on Cook's distance in our linear model*

| Predictor | Estimate | Pr(>|t|) | Predictor | Estimate | Pr(>|t|) |
|---|---|---|---|---|---|
| **mosque** | 0.025610 | 0.000000 | florist | 0.000104 | 0.000000 |
| **roofing contractor** | 0.004506 | 0.000000 | transit station | 0.000079 | 0.000000 |
| **airport** | 0.002418 | 0.000000 | liquor store | 0.000059 | 0.000000 |
| **taxi stand** | 0.002145 | 0.000000 | shoe store | 0.000057 | 0.000000 |
| **electrician** | 0.001384 | 0.000000 | car rental | 0.000044 | 0.000050 |
| **amusement park** | 0.001215 | 0.000000 | parking | 0.000031 | 0.000389 |
| **plumber** | 0.000853 | 0.000000 | casino | 0.000029 | 0.000001 |
| **Hindu temple** | 0.000850 | 0.000000 | shopping mall | 0.000025 | 0.000000 |
| **bowling alley** | 0.000413 | 0.000000 | accounting | 0.000025 | 0.000000 |
| **bicycle store** | 0.000399 | 0.000000 | hardware store | 0.000021 | 0.000000 |
| **cemetery** | 0.000393 | 0.000000 | bar | 0.000019 | 0.000000 |
| **locksmith** | 0.000372 | 0.000000 | travel agency | 0.000017 | 0.000000 |
| **storage** | 0.000328 | 0.000000 | university | 0.000016 | 0.000000 |
| **veterinary care** | 0.000302 | 0.000000 | department store | 0.000016 | 0.000001 |
| **local government office** | 0.000175 | 0.000000 | clothing store | 0.000015 | 0.000000 |
| **museum** | 0.000162 | 0.000000 | hair care | 0.000014 | 0.000034 |
| **police** | 0.000117 | 0.000000 | home goods store | 0.000011 | 0.000000 |
| **place of worship** | 0.000111 | 0.000000 | car repair | 0.000008 | 0.000000 |
| **fire station** | 0.000109 | 0.000000 | | | |

$H_5$ (+Machine Learning ) : $H_{3e}$ (+Machine Learning ) Feature selection to maximize AIC resulted in a very modest 4% decrease in mean squared error. More importantly, despite throwing a huge number of computer cycles at the problem, the machine learning approach did not perform any better than a human could by simply removing the predictors with the lowest p-values from the regression.

## 6. Confounding Effects

In this section, we provide a test to eliminate confounding effects of sub-optimal revenues arising from "best practices" credit scoring algorithms denying credit to otherwise good customers who would repay their loans; or sub-optimal costs arising from providing credit to borrowers who will ultimately default on those loans.

$H_6$: The company's P2P loan profitability business model is independent of the structure of the social network in which borrowers communicate with each other

The test is relatively straightforward. We reshaped the dataset into a directed graph, computed the graph statistics and then we loaded these into a table of senders and receivers of voice telephone calls and SMSs. We regressed these against an effect indicator variable for default. The value $R^2 \cong 9.6\%$ is relatively low at <10% but this is not unusual for credit scoring algorithms where any additional variance explained can result in additional profitability. Investment models, credit scoring models and racetrack betting models often have fits where $R^2$ is less than 10% and even less than 1%. Over repeated use these small amounts eventually add to increased profitability.

*Table 9: default ~ graph statistics regression*

|  | Estimate | Std. | t | Pr(>|t|) |  |
|---|---|---|---|---|---|
| *(Intercept)* | -0.04097 | 0.00410 | -10.00500 | 0.00000 | *** |
| *Number of out edges* | -0.07558 | 0.00331 | -22.83200 | 0.00000 | *** |
| *Number of in edges* | -0.09644 | 0.00196 | -49.34000 | 0.00000 | *** |
| *The number of triads that the node participates in* | 0.00262 | 0.00005 | 54.11900 | 0.00000 | *** |
| *Total communication duration* | 0.00000 | 0.00000 | -0.06800 | 0.94600 | |
| *Amount of loan* | 0.00001 | 0.00000 | 73.67600 | 0.00000 | *** |
| *Loan interest rate* | 0.03287 | 0.00062 | 52.85600 | 0.00000 | *** |
| *Eigenvector centrality* | 8489.00000 | 372.40000 | 22.79300 | 0.00000 | *** |
| *Farness centrality* | 0.00000 | 0.00000 | 1.32900 | 0.18400 | |

$R^2$ 0.09585, F-statistic 1610 on 8 and 121,383 DF

Regression statistics confirm some of our intuition from looking at the graph in figures 1 and 2. Good borrowers tend to be better connected to the rest of the graph than are borrowers who are likely to default. As the number of *out* edges increases for the average borrower node, the default rate drops by 7.6% for each additional out edge; for each additional *in* edge it drops by 9.6%. The average borrower node participates in 4.5 triads, but increase participation I hate single triad only increases default rate by 1/5 of 1%. The loan amount and the loan interest rate are strongly correlated to default, which makes sense because credit scoring is going to allocate smaller loans and higher interest rates to borrowers who are likely to default. Eigenvector centrality is a measure of the influence of a node in a network (Google's PageRank and the Katz centrality are variants of eigenvector centrality). Eigenvectors centralities tend to be small numbers so the coefficient is quite large; loan defaulters appear to be much more tightly clustered around each other, a situation which we could infer from figures 2 and 3, with a aggregate into small clusters. Farness centrality similarly measures influence, and is the average shortest path length to all other nodes that reside in the same connected component as the given node. It is easier to compute that eigenvector centrality but the regression statistics show it contains less information about default.

## 7. Discussions and Conclusions

This research investigated the potential for improving P2P credit scoring by using the private information about communications at location visits of borrowers. We initiated the research by looking at actual and potential structural models P2P of borrower communications and travels.

$H_6$ investigated whether P2P borrowers' ego networks exhibit scale-free behavior driven by underlying preferential attachment mechanisms that connect borrowers in a fashion that potentially could be used to predict credit defaults. Our analysis weeklies supported please feel free assumption with exponent ~3. Perturbing the data sets by removing nodes from the dataset caused this exponent to vary somewhat between 2.5 and 6. This was thought to be partly the byproduct of our data set having only 784 loans, and being somewhat disconnected. We surmise that with larger data samples, scale free behavior will be more robust, and consequently graph metrics are expected to be more informative in assessing the profitability of a loan.

Next, we analyzed the confounding effects of higher costs arising from "best practices" credit scoring algorithms denying credit to otherwise good customers would repay their loans; or higher revenues arising from providing credit to borrowers who will ultimately default on those loans. We also found that a company's P2P loan profitability business model is independent of the structure of the social network in which borrowers communicate with each other and that's that statistics summarizing these communications can be used as an independent source of information for a credit risk analysis.

Graph topology was found to be an important predictor of loan profitability, explaining over 5 ½% of variability in our synthetic profitability measure, or $22,615.27 of the total profit (synthetic) of $411,186.80 for the 784 loans in our dataset. Our tests of $H_3$ indicates that the greater centrality of a particular borrower in the communication network is correlated strongly with loan profitability; while high numbers of outgoing edges (communications to others) correlate strongly to losses and defaults.

Borrower location data was similarly found to be an important predictor of loan. Although any observations are inherently subjective, we can surmise that a proximity of certain types of businesses reflects the borrowers' particular lifestyle choices, and these lifestyle choices impact ability to repay loans on time. Locations such as dentist offices, pharmacies, veterinary care facilities, libraries and so forth reflect lifestyles of borrowers who take responsibility for their homes and families pass and probably loans.   Locations such as nightclubs, bus stations, City Hall, airports and financial offices may reflect lifestyles of people who are more carefree, and perhaps more likely to get into financial trouble.   Location information about types of businesses or institutions that where we been 50 m of the borrower at the time of communication explain an additional 19% of the synthetic profitability variance, or $77,632.07 of the total profit (synthetic) of $411,186.80 for the 784 loans in our dataset.

Machine learning proved to be less useful, at least partly because it revealed few things that linear regression hadn't revealed previously.  We rejected $H_5$    Feature selection to maximize AIC resulted in only a modest 4% decrease in mean squared error.  More importantly, despite throwing a huge number of compute cycles at the problem, the machine learning approach did not perform any better than a human could by simply removing the predictors with the lowest p-values from the regression.

## 8. References

Akaike, H. (1973). Information theory and an extension of the maximum lilelihood principle. 2nd International Symposium on Information Theory, Akademiai Kiado, Budapest, 1973.

Arnaboldi, V., et al. (2012). Analysis of ego network structure in online social networks. Privacy, security, risk and trust (PASSAT), 2012 international conference on and 2012 international confernece on social computing (SocialCom), IEEE.

Böhme, R. and S. Pötzsch (2010). Privacy in Online Social Lending. AAAI Spring Symposium: Intelligent Information Privacy Management.

Borgatti, S. P., et al. (2009). "Network analysis in the social sciences." Science **323**(5916): 892-895.

Burnham, K. P. and D. R. Anderson (2003). Model selection and multimodel inference: a practical information-theoretic approach, Springer Science & Business Media.

Callaway, D. S., et al. (2000). "Network robustness and fragility: Percolation on random graphs." Physical review letters **85**(25): 5468.

Cook, R. D. (1977). "Detection of influential observation in linear regression." Technometrics **19**(1): 15-18.

Cook, R. D. (1979). "Influential observations in linear regression." Journal of the American Statistical Association **74**(365): 169-174.

Csardi, G. and T. Nepusz (2006). "The igraph software package for complex network research." InterJournal, Complex Systems **1695**(5): 1-9.

Csárdi, G. and T. Nepusz (2010). "igraph Reference Manual." URL: http://igraph/. sourceforge. net/documentation. html (accessed April **20**.

de Nooy, W. (2012). "Graph theoretical approaches to social network analysis." Computational complexity: theory, techniques, and applications. Springer, Heidelberg: 2864-2877.

Dillon, T. W. and D. Lending (2010). "Will they adopt? Effects of privacy and accuracy." Journal of Computer Information Systems **50**(4): 20-29.

Easley, D. and J. Kleinberg (2010). Networks, crowds, and markets: Reasoning about a highly connected world, Cambridge University Press.

Economist (2017). "In fintech, China shows the way."

Everett, M. and S. P. Borgatti (2005). "Ego network betweenness." Social networks **27**(1): 31-38.

Fang, Y. (2011). "Asymptotic equivalence between cross-validations and Akaike information criteria in mixed-effects models." Journal of Data Science **9**(1): 15-21.

Grodzinsky, F. S. and H. T. Tavani (2005). "P2P networks and the Verizon v. RIAA case: implications for personal privacy and intellectual property." Ethics and information technology **7**(4): 243-250.

Jones, C. and E. H. Volpe (2011). "Organizational identification: Extending our understanding of social identities through social networks." Journal of organizational Behavior **32**(3): 413-434.

Kadushin, C. (2012). Understanding social networks: Theories, concepts, and findings, OUP USA.

Leskovec, J. and J. J. Mcauley (2012). Learning to discover social circles in ego networks. Advances in neural information processing systems.

Leskovec, J. and R. Sosič (2016). "Snap: A general-purpose network analysis and graph-mining library." ACM Transactions on Intelligent Systems and Technology (TIST) **8**(1): 1.

Scott, J. (2017). Social network analysis, Sage.

Scott, W. R. and G. F. Davis (2003). "Networks in and around organizations." Organizations and Organizing.

Strogatz, S. H. (2001). "Exploring complex networks." Nature **410**(6825): 268.

Wasserman, S. and K. Faust (1994). Social network analysis: Methods and applications, Cambridge university press.